%% file: fpcp03pre.tex
\documentclass[12pt]{article}
\usepackage{epsfig,graphicx}
\usepackage{fpcp03}

\input fpcp03-macros
\begin{document}

\begin{flushright}  
UdeM-GPP-TH-03-110\\
\end{flushright}
\vskip2truemm

\Title{New Physics and $B\to V_1 V_2$ Decays\footnote{Talk given at
{\it Flavor Physics and CP Violation (FPCP2003)}, \'Ecole
Polytechnique, Paris, France, June 2003.}}

\bigskip

\label{LondonStart}

\author{David London\index{London, D.}}

\address{
Laboratoire Ren\'e J.-A. L\'evesque, Universit\'e de Montr\'eal, \\
C.P. 6128, succ.~centre-ville, Montr\'eal, QC, Canada H3C 3J7 \\
}

\makeauthor\abstracts{I discuss several techniques for observing new
physics using measurements of $B\to V_1 V_2$ decays. Within the
standard model, all CP-violating triple-product correlations (TP's)
involving light vector mesons are expected be very small or to
vanish. However, these TP's can be large in models with new physics
(NP).  If a time-dependent angular analysis of $B\to V_1 V_2$ can be
performed, there are numerous additional tests for NP in the decay
amplitudes. Should a signal for NP be found, one can place constraints
on the NP parameters.}

\section{Triple Products}

It is well known that $B\to V_1 V_2$ decays cannot be simply used to
measure indirect CP-violating asymmetries and obtain clean information
about weak phases. The reason is that the state $V_1 V_2$ is not a CP
eigenstate -- it involves three helicity amplitudes. Two of these
($A_0$, $A_\|$) are CP-even, while the third ($A_\perp$) is CP-odd. As
a function of these helicity amplitudes, the $B\to V_1 V_2$ decay
amplitude can be written as \cite{DDLR}
\begin{equation}
M = A_0 \varepsilon_1^{* L} \cdot \varepsilon_2^{* L} 
- {1 \over \sqrt{2}} A_\| {\vec\varepsilon}_1^{* T} \cdot
  {\vec\varepsilon}_2^{* T}
- {i \over \sqrt{2}} A_\perp {\vec\varepsilon}_1^{* T} \times
  {\vec\varepsilon}_2^{* T} \cdot {\hat p} ~,
\end{equation}
where ${\hat p}$ is the unit vector along the direction of motion of
$V_2$ in the rest frame of $V_1$, and $\varepsilon_{1,2}$ are
polarizations of vector mesons. In the above, $\varepsilon_i^{* L} =
{\vec\varepsilon}_i^* \cdot {\hat p}$, and ${\vec\varepsilon}_i^{* T}
= {\vec\varepsilon}_i^* - \varepsilon_i^{* L} {\hat p}$. On the other
hand, it is also well known that one can separate the helicity
amplitudes using a (time-dependent) angular analysis. In this way one
can measure the indirect CP asymmetries in each individual helicity
state.

However, this angular analysis contains a great deal more information,
due to the interference of CP-even and CP-odd amplitudes. The
time-integrated differential decay rate contains 6 angular terms. Two
of these are \cite{DDLR}
\begin{equation}
- {{\rm Im}(A_\perp A_0^*) \over 2\sqrt{2}} \sin 2\theta_1 \sin
2\theta_2 \sin\phi - {{\rm Im}(A_\perp A_\|^*) \over 2} \sin^2\theta_1
\sin^2\theta_2 \sin 2\phi ~.
\label{angdist}
\end{equation}
We have assumed that both vector mesons decay into pseudoscalars,
i.e.\ $V_1 \to P_1 P_1'$, $V_2 \to P_2 P_2'$. In the above, $\theta_1$
($\theta_2$) is the angle between the directions of motion of the
$P_1$ ($P_2$) in the $V_1$ ($V_2$) rest frame and the $V_1$ ($V_2$) in
the $B$ rest frame, and $\phi$ is the angle between the normals to the
planes defined by $P_1 P_1'$ and $P_2 P_2'$ in the $B$ rest frame. The
key point is that these two terms involve the {\it triple product}
${\vec\varepsilon}_1^{* T} \times {\vec\varepsilon}_2^{* T} \cdot
{\hat p}$. Triple products (TP's) are odd under time reversal (T) and
hence, by the CPT theorem, also constitute potential signals of CP
violation. (Note that a full angular analysis is not necessary to
measure TP's.)

Now, it is well-known that triple product signals are not necessarily
CP-violating -- even if the weak phases vanish, nonzero TP signals can
be produced by strong phases. In order to obtain a true CP-violating
signal, it is necessary to compare the TP in $B \to V_1 V_2$ with that
in ${\bar B} \to {\bar V}_1 {\bar V}_2$ \cite{Valencia:1988it}. Note
that the TP signal ${\vec\varepsilon}_1^{* T} \times
{\vec\varepsilon}_2^{* T} \cdot {\hat p}$ is odd under P. As a
consequence, the true CP-violating triple product is found by {\it
adding} the two T-odd asymmetries:
\begin{equation}
{\cal A}_T \equiv {1\over 2}(A_T + {\bar A}_T) ~.
\end{equation}
This is an important point: since one has to add $A_T$ and ${\bar
A}_T$, neither tagging nor time dependence is necessary to measure
TP's! In principle, one can combine measurements of charged and
neutral $B$ decays to obtain a triple-product signal.

Triple products are of particular interest because they are
complementary to direct CP asymmetries. Both signals can be nonzero
only if there are two interfering decay amplitudes in a given $B$
decay. However, denoting $\phi$ and $\delta$ as the relative weak and
strong phases, respectively, between the two interfering amplitudes,
the expressions for the two signals can be written
\begin{equation}
{\cal A}_{CP}^{dir} \propto \sin\phi \sin\delta  ~~,~~~~
{\cal A}_T \propto \sin\phi \cos\delta ~. 
\end{equation}
If the strong phases are small, as may well be the case in $B$ decays,
all direct CP-violation signals will be tiny as well. On the other
hand, TP asymmetries are {\it maximal} when the strong-phase
difference vanishes. Thus, it may well be more promising to search for
triple-product asymmetries than direct CP asymmetries in $B$ decays.

\section{Triple Products in the Standard Model}

As mentioned above, all CP-violating effects require the interference
of two amplitudes, with different weak phases. Certain decays in the
standard model (SM), such as those dominated by $b\to c {\bar c} s$ or
$b\to s {\bar s} s$, do not satisfy this. Thus, no triple products are
expected in $B \to J/\psi K^*$, $B \to \phi K^*$, $B\to D_s^* D^*$,
etc. However, other processes ($b\to u{\bar u} s$, $b\to c{\bar c} d$,
etc.) receive both tree and penguin contributions. Thus, it may be
possible to produce TP's in decays such as $B\to D^* {\bar D}^*$, $B
\to \rho K^*$, etc.\ \cite{Datta:2003mj}.

Consider now $B\to V_1 V_2$ decays within factorization. The amplitude
is
\begin{equation}
\sum_{{\cal O},{\cal O}'} \left\{ \bra{V_1} {\cal O} \ket{0} \bra{V_2}
{\cal O}' \ket{B} + \bra{V_2} {\cal O} \ket{0} \bra{V_1} {\cal O}'
\ket{B} \right\} ~.
\end{equation}
Note that TP's are a {\it kinematical} CP-violating effect -- unless a
given decay includes both of the above amplitudes, with a relative
weak phase, no TP will be produced. For example, even though the decay
$B_d^0 \to D^{*+} D^{*-}$ receives both a tree ($V_{cb}^* V_{cd}$) and
a penguin ($V_{tb}^* V_{td}$) contribution, there is no TP. The point
is that both of these amplitudes contribute to $\bra{D^{*+}} {\cal O}
\ket{0} \bra{D^{*-}} {\cal O}' \ket{B}$; there is no $\bra{D^{*-}}
{\cal O} \ket{0} \bra{D^{*+}} {\cal O}' \ket{B}$
amplitude. (Equivalently, in the SM one has only ${\bar b} \to {\bar
c}$ transitions; ${\bar b} \to c$ does not occur.) Thus, in the SM, no
TP is predicted in $B_d^0 \to D^{*+} D^{*-}$.

This then begs the question: which $B\to V_1 V_2$ decays are expected
to yield large triple products in the SM? The answer is simple: {\it
NONE} \cite{Datta:2003mj}. 

This can be understood via the following points:
\begin{enumerate}

\item If $V_1 = V_2$, no TP is possible, since there is only a single
  kinematical amplitude. Therefore, if $V_1$ and $V_2$ are related by
  a symmetry [e.g.\ isospin, flavour $SU(3)$], the TP is suppressed by
  the size of symmetry breaking.

\item The longitudinal amplitude $A_0$ is much larger than the
  transverse amplitudes $A_{\|,\perp}$. Therefore, TP's are suppressed
  by at least one power of $m_V/m_B$.

\item The interfering amplitudes are typically different in size,
  leading to further suppression of TP's.

\end{enumerate}
All of these factors lead to the suppression of TP's. The net effect
is that all TP's involving light vector mesons are either expected to
vanish or be very small in the SM. Note also that nonfactorizable
effects do not change this conclusion \cite{Datta:2003mj}.

Since all TP's in $B\to V_1 V_2$ decays with light vector mesons are
expected to be small, this is an excellent place to search for new
physics (NP). For example, at present the indirect CP asymmetry in
$B_d^0(t) \to \phi K_S$ differs from that in $B_d^0(t) \to J/\psi K_S$
\cite{phiKs}. If this discrepancy holds up --- it is not yet
statistically significant --- it would require a NP amplitude in
$B_d^0 \to \phi K_S$ \cite{Gross}. If present, this new amplitude
would also contribute to $B \to \phi K^*$, leading to triple products
in this decay. One of the many possibilities for this new physics is
supersymmetry with R-parity violation \cite{dattarparity}. Although
the TP in $B \to \phi K^*$ vanishes in the SM, we find that one can
get large TP asymmetries, in the range 15--20\%, in this model
\cite{Datta:2003mj}! This shows quite clearly that triple products are
an excellent way to search for physics beyond the SM.

\section{Time-Dependent Angular Analysis}

Consider now a $B$ decay which in the SM is dominated by a single
amplitude (e.g. $B \to J/\psi K$, $\phi K$, etc.). Suppose that there
is a new-physics amplitude, with a different weak phase, which
contributes to this decay. As I have argued above, such an amplitude
can be detected by looking for both direct CP violation and triple
products. However, as can be seen below, much more information can be
obtained if a time-dependent angular analysis of the corresponding
$B^0(t) \to V_1 V_2$ decay can be performed.

We write
\begin{eqnarray}
A_\lambda \equiv Amp (B \to V_1 V_2)_\lambda &=& a_\lambda e^{i
\delta_\lambda^a} + b_\lambda e^{i\phi} e^{i \delta_\lambda^b} ~, \\
{\bar A}_\lambda \equiv Amp ({\bar B} \to
{\overline{V}}_1 {\overline{V}}_2)_\lambda &=& a_\lambda e^{i
\delta_\lambda^a} + b_\lambda e^{-i\phi} e^{i \delta_\lambda^b} ~,
\end{eqnarray}
where $\lambda = \left\{0,\|,\perp \right\}$. The $a_\lambda$'s and
$b_\lambda$'s are the SM and NP amplitudes, respectively, $\phi$ is
the NP weak phase, and the $\delta_\lambda^{a,b}$ are the strong
phases. The time-dependent decay rate is given by
\begin{equation}
\Gamma(
{\hbox{$B$\kern-1.1em\raise1.4ex\hbox{{{\raise.35ex\hbox
{${\scriptscriptstyle (}$}}---{\raise.35ex\hbox{${\scriptscriptstyle )}$}}}}}}(t)
\to V_1 V_2) = e^{-\Gamma t} \sum_{\lambda\leq\sigma} \left(
\Lambda_{\lambda\sigma} \pm \Sigma_{\lambda\sigma}\cos(\Delta M t) \mp
\rho_{\lambda\sigma}\sin(\Delta M t) \right) g_\lambda g_\sigma ~,
\label{timedepdecayrate}
\end{equation}
where the $g_\lambda$ are known functions of the kinematic angles
$\theta_1$, $\theta_2$, $\phi$. There are 18 observables, all
functions of the $A_\lambda$ and ${\bar A}_\lambda$. For example,
\begin{eqnarray} 
& \Lambda_{\lambda\lambda} = \frac{1}{2}(|A_\lambda|^2+|{\bar
A}_\lambda|^2) ~~,~~~~
\Sigma_{\lambda\lambda} = \frac{1}{2}(|A_\lambda|^2-|{\bar
A}_\lambda|^2) ~, & \\
& \Lambda_{\perp i} = - {\rm Im} \left( A_\perp { A}_i^* - {\bar
A}_\perp {{\bar A}_i}^* \right) ~~,~~~~ \rho_{\lambda\lambda} =
\mp{\rm Im} \left( \frac{q}{p} A_\lambda^* {\bar A}_\lambda \right)
~. &
\end{eqnarray}
For a given helicity $\lambda$, $\Lambda_{\lambda\lambda}$ essentially
measures the total rate, while $\Sigma_{\lambda\lambda}$ and
$\rho_{\lambda\lambda}$ represent the direct and indirect CP
asymmetries, respectively. The quantity $\Lambda_{\perp i}$
($i=\{0,\|\}$) is simply the triple product discussed earlier.

Now, if there is no new physics (i.e.\ $b_\lambda = 0$), there are
only 6 theoretical parameters: 3 $a_\lambda$'s, 2 strong phase
differences, and the phase of $B^0$--${\bar B}^0$ mixing ($q/p$). This
implies that there are 12 relations among the observables:
\begin{eqnarray}
& \Sigma_{\lambda\lambda}= \Lambda_{\perp i}= \Sigma_{\| 0}=0 ~~,~~~~
\rho_{ii} / \Lambda_{ii} = - \rho_{\perp\perp} / \Lambda_{\perp\perp}
= \rho_{\|0} / \Lambda_{\| 0} ~, & \nonumber\\
& 2 \Lambda_{\|0} \Lambda_{\perp\perp} \left(
\Lambda_{\lambda\lambda}^2 -\rho_{\lambda\lambda}^2 \right)
= \left[ \Lambda_{\lambda\lambda}^2 \rho_{\perp 0} \rho_{\perp\|} +
  \Sigma_{\perp 0} \Sigma_{\perp \|} \left( \Lambda_{\lambda\lambda}^2
  -\rho_{\lambda\lambda}^2 \right) \right] ~, & \nonumber\\
& \rho_{\perp i}^2 \Lambda_{\perp\perp}^2 = \left(
\Lambda_{\perp\perp}^2 -\rho_{\perp\perp}^2 \right) \left(
4\Lambda_{\perp\perp} \Lambda_{ii}-\Sigma_{\perp i}^2 \right) ~. &
\end{eqnarray}
The violation of any of these relations will be a smoking-gun signal
of NP \cite{London:2003rk}. There are thus many more ways to search
for new physics if a time-dependent angular analysis can be done.

But there's more! Suppose that a signal for new physics is found,
implying that $b_\lambda \ne 0$. In this case there are 13 theoretical
parameters: 3 $a_\lambda$'s, 3 $b_\lambda$'s, 5 strong phase
differences, and two weak phases ($\phi$ and $q/p$). However, at best
one can measure the magnitudes and relative phases of the 6 decay
amplitudes $A_\lambda$ and ${\bar A}_\lambda$. That is, there are
really only 11 independent observables in Eq.~\ref{timedepdecayrate}.
Naively, one would imagine that, with 11 measurements and 13 unknowns,
one cannot get any information about the new physics, even if there is
a NP signal. However, this is not correct: because the expressions
relating the observables to the theoretical parameters are nonlinear,
one can actually {\it constrain} the NP parameters \cite{London:2003rk}.

For example, if $\Sigma_{\lambda\lambda} \ne 0$, 
\begin{equation}
b^2_\lambda \ge {1\over 2} \Lambda_{\lambda\lambda} \left[ 1 -
\sqrt{1 - \Sigma_{\lambda\lambda}^2/ \Lambda_{\lambda\lambda}^2}
\right].
\end{equation}
Similarly, if $\Sigma_{\lambda\lambda} = 0$, but $\Lambda_{\perp i}
\ne 0$,
\begin{equation}
2 (b_i^2 \mp b_\perp^2) \geq \Lambda_{ii} \mp \Lambda_{\perp\perp} -
\sqrt{ \left( \Lambda_{ii} \mp \Lambda_{\perp\perp}\right)^2 \pm
\Lambda_{\perp i}^2} ~.
\end{equation}

Also
\begin{eqnarray}
\Lambda_{ii} \cos\eta_i + \Lambda_{\perp\perp} \cos(\eta_\perp - 2
\eta_i) & \le & \sqrt{ \left( \Lambda_{ii} + \Lambda_{\perp\perp}
\right)^2 - \Lambda_{\perp i}^2}~, \nonumber\\
\Lambda_{ii} \cos\eta_i - \Lambda_{\perp\perp} \cos \eta_\perp
& \le & \sqrt{ \left( \Lambda_{ii} - \Lambda_{\perp\perp}
\right)^2 + \Lambda_{\perp i}^2}~,
\end{eqnarray}
where $\eta_\lambda \equiv 2 \left( {q\over p}^{\mathit{meas}}_\lambda
- {q\over p}^{mix} \right)$. If $\Lambda_{\perp i} \ne 0$, one cannot
have $\eta_i = \eta_\perp = 0$. Thus, one obtains a lower bound on the
difference between the measured value and the true value of the phase
of $B^0$--${\bar B}^0$ mixing.

\section{Conclusion}

$B \to V_1 V_2$ decays contain an enormous amount of information,
especially if an angular analysis can be performed. One very useful
class of measurements is triple-product correlations (TP's):
${\vec\varepsilon}_1^{* T} \times {\vec\varepsilon}_2^{* T} \cdot
{\hat p}$. A true CP-violating triple-product signal can be obtained
by {\it adding} the TP's found in $B \to V_1 V_2$ and ${\bar B} \to
{\bar V}_1 {\bar V}_2$. Thus, neither tagging nor time dependence is
necessary to measure TP's -- in principle, the measurements of charged
and neutral $B$ decays can be combined. 

We have examined the size of TP's in the standard model (SM). We find
that all TP's involving light vector mesons are either expected to
vanish or be very small. This makes triple products an excellent place
to search for new physics. Indeed, we have found that TP's which
vanish in the SM can be large (15--20\%) in the presence of new
physics.

If a full time-dependent angular analysis can be performed, much more
information is available. First, there are many more signals of new
physics. And second, should a signal for new physics be found, one can
place a lower limit on the size of the new-physics amplitudes, as well
as on their effect on the measurement of the phase of $B^0$--${\bar
B}^0$ mixing.

\section*{Acknowledgments}

I thank A. Datta, N. Sinha and R. Sinha for collaborations on the
topics discussed here. This work was financially supported by NSERC of
Canada.

%
\label{LondonEnd}
 
\end{document}

%% file: fpcp03-macros.tex
\def\beq{\begin{equation}}
\def\eeq#1{\label{#1}\end{equation}}
\def\eeqn{\end{equation}}

\def\beqa{\begin{eqnarray}}
\def\eeqa#1{\label{#1}\end{eqnarray}}
\def\eeqan{\end{eqnarray}}

\let\bar=\overbar

\def\bra#1{\left\langle{ #1} \right|}
\def\ket#1{\left| {#1} \right\rangle}

\def\Dslash{\not{\hbox{\kern-4pt $D$}}}
\def\dslash{\not{\hbox{\kern-2pt $\del$}}}

\def\msb{{\bar{\ssstyle M \kern -1pt S}}}

\def\BB0bar{B^0 {\overline B}^0}
\def\BB0dbar{B_d^0 {\overline B}_d^0}
\def\BB0sbar{B_s^0 {\overline B}_s^0}

\RequirePackage{xspace}

\usepackage{relsize}
\def\babar{\mbox{\slshape B\kern-0.1em{\smaller A}\kern-0.1em
    B\kern-0.1em{\smaller A\kern-0.2em R}}}

\def\Kbar  {\kern 0.2em\overline{\kern -0.2em K}{}\xspace}

\def\Kz    {\ensuremath{K^0}\xspace}
\def\Kzb   {\ensuremath{\Kbar^0}\xspace}
\def\KzKzb {\ensuremath{\Kz \kern -0.16em \Kzb}\xspace}
\def\Kp    {\ensuremath{K^+}\xspace}
\def\Km    {\ensuremath{K^-}\xspace}

\def\KpKm  {\ensuremath{\Kp \kern -0.16em \Km}\xspace}

\def\Dbar    {\kern 0.2em\overline{\kern -0.2em D}{}\xspace}

\def\Dz      {\ensuremath{D^0}\xspace}
\def\Dzb     {\ensuremath{\Dbar^0}\xspace}
\def\DzDzb   {\ensuremath{\Dz {\kern -0.16em \Dzb}}\xspace}
\def\Dp      {\ensuremath{D^+}\xspace}
\def\Dm      {\ensuremath{D^-}\xspace}

\def\DpDm    {\ensuremath{\Dp {\kern -0.16em \Dm}}\xspace}

\def\Bbar    {\kern 0.18em\overline{\kern -0.18em B}{}\xspace}

\def\BB      {\ensuremath{B\Bbar}\xspace} 
\def\Bz      {\ensuremath{B^0}\xspace}
\def\Bzb     {\ensuremath{\Bbar^0}\xspace}
\def\BzBzb   {\ensuremath{\Bz {\kern -0.16em \Bzb}}\xspace}
\def\Bu      {\ensuremath{B^+}\xspace}
\def\Bub     {\ensuremath{B^-}\xspace}

\def\BpBm    {\ensuremath{\Bu {\kern -0.16em \Bub}}\xspace}

\mathchardef\Upsilon="7107
\def\Y#1S{\ensuremath{\Upsilon{(#1S)}}\xspace}

\mathchardef\Deltares="7101
\mathchardef\Xi="7104
\mathchardef\Lambda="7103
\mathchardef\Sigma="7106
\mathchardef\Omega="710A

\def\Deltabar{\kern 0.25em\overline{\kern -0.25em \Deltares}{}\xspace}
\def\Lbar{\kern 0.2em\overline{\kern -0.2em\Lambda\kern 0.05em}\kern-0.05em{}\xspace}
\def\Sigbar{\kern 0.2em\overline{\kern -0.2em \Sigma}{}\xspace}
\def\Xibar{\kern 0.2em\overline{\kern -0.2em \Xi}{}\xspace}
\def\Obar{\kern 0.2em\overline{\kern -0.2em \Omega}{}\xspace}
\def\Nbar{\kern 0.2em\overline{\kern -0.2em N}{}\xspace}
\def\Xb{\kern 0.2em\overline{\kern -0.2em X}{}\xspace}

\newcommand{\tev}{\ensuremath{\mathrm{\,Te\kern -0.1em V}}\xspace}
\newcommand{\gev}{\ensuremath{\mathrm{\,Ge\kern -0.1em V}}\xspace}
\newcommand{\mev}{\ensuremath{\mathrm{\,Me\kern -0.1em V}}\xspace}
\newcommand{\kev}{\ensuremath{\mathrm{\,ke\kern -0.1em V}}\xspace}
\newcommand{\ev}{\ensuremath{\mathrm{\,e\kern -0.1em V}}\xspace}
\newcommand{\gevc}{\ensuremath{{\mathrm{\,Ge\kern -0.1em V\!/}c}}\xspace}
\newcommand{\mevc}{\ensuremath{{\mathrm{\,Me\kern -0.1em V\!/}c}}\xspace}
\newcommand{\gevcc}{\ensuremath{{\mathrm{\,Ge\kern -0.1em V\!/}c^2}}\xspace}
\newcommand{\mevcc}{\ensuremath{{\mathrm{\,Me\kern -0.1em V\!/}c^2}}\xspace}

\def\mus  {\ensuremath{\rm \,\mus}\xspace}

\def\mus        {\ensuremath{\,\mu{\rm s}}\xspace}    

\def\to                 {\ensuremath{\rightarrow}\xspace}

\def\pep2{PEP-II}

\def\gsim{{~\raise.15em\hbox{$>$}\kern-.85em
          \lower.35em\hbox{$\sim$}~}\xspace}
\def\lsim{{~\raise.15em\hbox{$<$}\kern-.85em
          \lower.35em\hbox{$\sim$}~}\xspace}

\xspace

\def\jetset74   {\mbox{\tt Jetset \hspace{-0.5em}7.\hspace{-0.2em}4}\xspace}
